\documentclass[aip,apl,reprint,floatfix,superscriptaddress]{revtex4-1}

\usepackage{epsf}
\usepackage{epsfig}
\usepackage{graphicx}
\usepackage{dcolumn}
\usepackage{bm}
\usepackage{amsfonts}
\usepackage{amsmath}

\usepackage{amssymb}
\usepackage{color,soul}

\input epsf

\begin{document}

\title{Quantum charge pumping in graphene-based devices: When lattice defects do help}

\author{Lucas H. Ingaramo}
\affiliation{Facultad de Matem\'atica, Astronom\'{\i}a y F\'{\i}sica (FaMAF), Universidad Nacional de C\'{o}rdoba, Ciudad Universitaria, 5000 C\'{o}rdoba, Argentina.}
\author{Luis E. F. Foa Torres}
\affiliation{Facultad de Matem\'atica, Astronom\'{\i}a y F\'{\i}sica (FaMAF), Universidad Nacional de C\'{o}rdoba, Ciudad Universitaria, 5000 C\'{o}rdoba, Argentina.}
\affiliation{Instituto de F\'{\i}sica Enrique Gaviola (IFEG), CONICET and
FaMAF, Universidad Nacional de C\'{o}rdoba, Ciudad Universitaria, 5000
C\'{o}rdoba, Argentina.}

\begin{abstract}
\begin{flushleft}\end{flushleft}
Quantum charge pumping, the quantum coherent generation of a dc current at zero bias through time-dependent potentials, provides outstanding opportunities for metrology and the development of nanodevices. The long electronic coherence times and high quality of the crystal structure of graphene may provide suitable building blocks for such quantum pumps. Here, we focus in adiabatic quantum pumping through graphene nanoribbons in the Fabry-P\'erot regime highlighting the crucial role of defects by using atomistic simulations. We show that even a single defect added to the pristine structure may produce a two orders of magnitude increase in the pumped charge.
\end{abstract}

\maketitle

Graphene and carbon nanotubes exhibit unprecedented electronic mean-free paths of up to several microns \cite{Bolotin2008,Mayorov2011}. Their high quality (defect-free) crystalline structure together with the ability of building low resistance contacts is evidenced in the observation of high conductances approaching the quantum limit \cite{Javey2004,Miao2007}.
Fabry-P{\'e}rot oscillations, a hallmark of ballistic transport, are experimentally observed in both carbon nanotubes \cite{Liang2001}, graphene \cite{Wu2012,Oksanen2013} and in careful experiments in suspended graphene \cite{Grushina2013,Schon2013}. After the initial excitement, the study of \textit{defected} samples came back to the main stage\cite{Gomez-Navarro2005,Krasheninnikov2007,Latil2004a,Avriller2006,Lherbier2008}: Defect engineering of carbon-based materials may provide alternative ways of tailoring nanomaterials \cite{Krasheninnikov2007,Terrones2010}.

Another issue of interest is the use of ac fields such as gate voltages or illumination with a laser to achieve unique phenomena in these low-dimensional materials such as laser-induced band gaps in graphene \cite{Oka2009,Calvo2011,Calvo2012a} which would allow for switching of the dc electrical response, or the generation of a dc current even in the absence of an applied bias voltage either by exploiting a ratchet effect \cite{Drexler2013} or quantum interference \cite{Connolly2013}. The latter phenomenon, known as quantum charge pumping \cite{Thouless1983,Altshuler1999,Buettiker2006}, is usually produced by modulating the sample properties through gate voltages which alternate with the same frequency but with a phase difference. The regime which is most usually explored corresponds to either isolated resonances or pristine, defect-free, materials.  Quantum pumping allows for the study of fundamental issues related to the breaking of symmetries in quantum transport while at the same time may provide for devices with lower power dissipation or even close the metrological triangle when pumping becomes quantized as in recent experiments\cite{Connolly2013}.  
Recent studies focused on the possible realization of this phenomenon in carbon-based devices, in both the adiabatic \cite{Prada2009,Zhu2009,Grichuk2010,Alos-Palop2011,Perroni2013} and non-adiabatic limits \cite{FoaTorres2011,San-Jose2011,Zhou2012}, but the influence defects, which is the subject of the present work, has not been addressed. 

\begin{figure}
\centering
\includegraphics[width=\columnwidth]{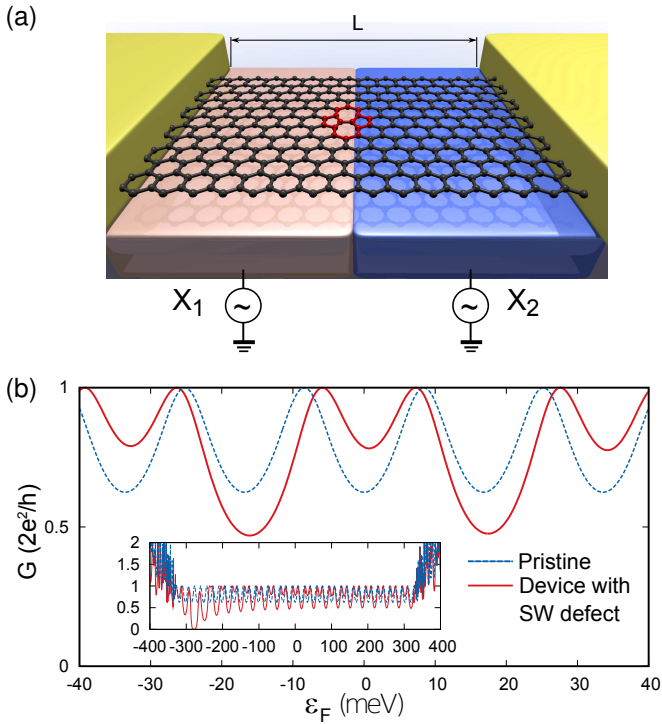}
\caption{(Color online) (a) Scheme of the considered setup where a graphene ribbon of lenght $L$ is connected to electrodes and two time-dependent gate voltages are applied ($X_1$ and $X_2$). (b) Conductance as a function of the Fermi energy for a ribbon of 111.4 nm length and 3 nm width with slightly imperfect contacts ($\gamma_{FP}=0.7 \gamma_0$) and $X_1=X_2=0$. The dashed line corresponds to a pristine sample while the solid one is for a device with one Stone-Wales defect. The inset shows a broader scale for the Fermi energy reaching the limits of the first conductance plateau.}
\label{fig1}
\end{figure}

Here we consider the paradigmatic case of a graphene nanoribbon in the Fabry-P\'erot regime with two out-of-phase alternating gate voltages as is usual for adiabatic pumping (see scheme in Fig.\ref{fig1}-a). The electrical response at zero dc bias is then solved by combining a semi-empirical Hamiltonian including a lattice defect with Green's functions in the adiabatic limit \cite{Brouwer1998}. While defects normally degrade the conductance and may even completely hinder charge transport, we show that defects may amplify the interferences which are at the heart of the pumping mechanism, thereby leading to a dramatic enhancement of the pumped current (of up to 2-3 orders of magnitude).

To start with we consider a standard $\pi$-orbitals Hamiltonian \cite{Charlier2007} for a (armchair) graphene nanoribbon:
\begin{equation}
{\cal H}_e=\sum_{i} E_i {\hat c}_i^{\dag} {\hat c}_i^{}-\sum_{\langle i,j \rangle} \gamma_{i,j} [ {\hat c}_i^{\dag} {\hat c}_j^{} + {\hat c}_j^{\dag}  {\hat c}_i^{} ]
\label{eq-H}
\end{equation}
where ${\hat c}_i^{\dag}$ and ${\hat c}_i^{}$ are the electronic creation and anihilation operators at site $i$, $E_i$ is the site energy and $\langle i,j \rangle$ denote that the summation is restricted to nearest neighbors. The transfer integral between nearest neighbors is chosen as $\gamma_0=2.7eV$ \cite{Charlier2007}. To simulate the effect of a partially transparent contact to semi-infinite graphene electrodes (left and right) the hopping matrix elements connecting a region of length $L$ of the device to the electrodes are affected by a factor $\gamma_{FP}<1$. We note that in this high-conductance regime transport is very close to the quantum limit for ballistic transport and therefore charging effects \cite{Beloborodov2007} do not play a role\cite{footnote} as evidenced by experiments \cite{Liang2001}. In our case, the gate voltages $X_1(t)$ and $X_2(t)$ simply shift the corresponding site energies in the same amount $E_{i\in j}=eX_j(t)$. We carried out simulations for different types of lattice defects with similar results. Here we concentrate in Stone-Wales topological defects. These defects are formed by rotating a carbon-carbon bond 90 degrees leading to a pentagon-heptagon pair (as represented in red in the scheme in Fig.\ref{fig1}-a) \cite{Matsumura2001,Romeo2011}.

In a situation such as the one schematically represented in Fig. \ref{fig1}-a, a dc current is generated as a result of the cyclic variation of two time-dependent gate voltages $(X_1(t), X_2(t))$. When the time-variation is slow enough, this pumped current can be calculated in an elegant way by using the adiabatic theory \cite{Brouwer1998}. The charge pumped per cycle is constant and can be written as an integral over the contour in the $X_1-X_2$ plane as\cite{Brouwer1998}:

\begin{equation}
	Q(m,r)=\frac{e}{\pi} \int_{A} dX_{1} dX_{2} \sum_{\beta} \sum_{\alpha \in m} \Im \frac{\partial S^{*}_{\alpha \beta}}{\partial X_{1}} \frac{\partial S_{\alpha \beta}}{\partial X_{2}}
	\label{pumping}
\end{equation}
where $S_{\alpha \beta}$ are the matrix elements of the scattering matrix encoding the probability amplitudes for the reflection/transmission between the differente electrodes, $\alpha,\beta=L,R$. One can also define the pumping kernel $dQ$ as 

\begin{equation}
	dQ=
	\sum_{\beta} \sum_{\alpha \in m}
	\Im 
	\frac{\partial S^{*}_{\alpha \beta}}{\partial X_{1}} 
	\frac{\partial S_{\alpha \beta}}{\partial X_{2}},
	\label{kernel}
\end{equation}
which has the advantage of being a contour-independent property.

\begin{figure}
\centering
\includegraphics[width=0.8\columnwidth]{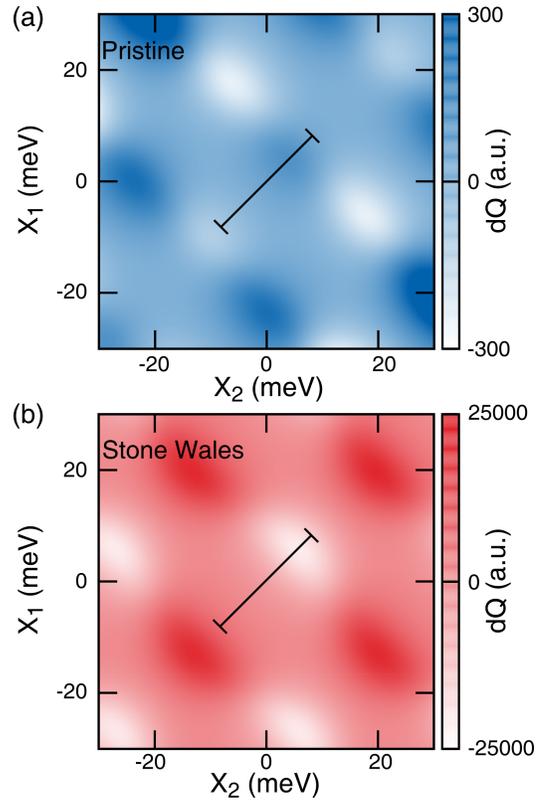}
\caption{(Color online) The charge pumping kernel $dQ$ is shown in color scale as a function of the gate voltages $X_1$ and $X_2$. (a) shows the case of a pristine armchair graphene nanoribbon of 3 nm width and 111.4 nm length in the Fabry-P\'erot regime. (b) shows the same for a device with a single Stone-Wales defect. Note the large change (almost three orders of magnitude) in the intensity of the pumped charge (color bars on the right). The black bar indicates the energy level spacing for this length ($16.4$ meV).}
\label{fig2}
\end{figure}

In absence of defects, the model presented above gives a series of conductance oscillations as shown in Fig. \ref{fig1}-b with a dashed line. These so-called Fabry-P\'erot oscillations were observed experimentally \cite{Liang2001,Oksanen2013} in devices with low-resistance contacts. Close to the charge neutrality point, the spacing between the maxima is approximately constant $\Delta\sim \hbar v_F/L $, $v_F$ is the Fermi velocity and $L$ is the device length. The amplitude of these oscillations is controlled by the matrix element $t_{FP}$ and an analytical solution in terms of Chebyshev polynomials is feasible \cite{Nemec2008}. When adding the two slowly varying gate voltages $X_1$ and $X_2$, a pumped charge is generated. This pumped charge follows the structure of maxima and minima observed in the conductance, see Fig. \ref{fig2}-a.

The addition of a single Stone-Wales defect introduces important depressions within the first conductance plateau, close to the onset of higher subbands \cite{Matsumura2001}. However, since these destructive interference features are far from the charge neutrality point (at least for small ribbons), they may be difficult to reach experimentally. A more detailed look to the results on Fig. \ref{fig1}-b (solid line) reveals that the defects also introduce smaller modifications of the conductance oscillations close to the Dirac point: the period of the oscillations changes due to the additional scatterer (which roughly divides the sample in two parts) and their amplitude increases on about 40\% per cent. Although this change in the conductance may seem relatively weak, it turns out that the pumped charge is being dramatically affected by the defect. Indeed, Fig. \ref{fig2}-b shows that, as compared to the pristine system (Fig. \ref{fig2}-a), the pumped charge for the defective system increases in almost three orders of magnitude.

\begin{figure}
	\centering
	\includegraphics[width=\columnwidth]{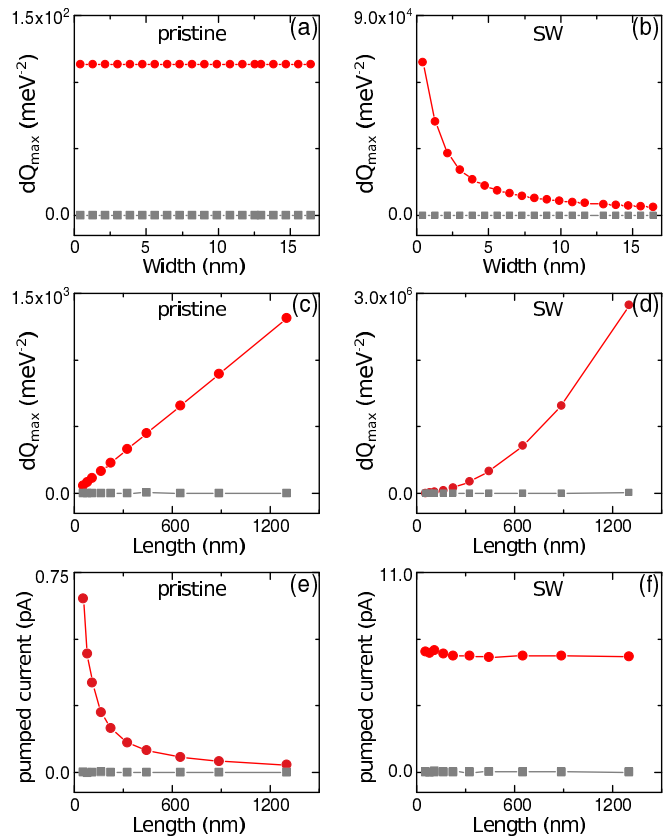}
	\caption{(Color online) Maxima of the pumped charge kernel ($dQ^{max}$) (red dots) as a function of the device width for (a) pristine system and one with a Stone-Wales (SW) defect. The contribution due to the transmission channel is shown with grey squares to emphasize that the dominant contribution comes from the reflection amplitude. The device length in this case is 111.4 nm and the Fermi energy is chosen to be $\varepsilon=13.5$meV. (c) and (d) show the scaling with the device length for the same cases. (e) and (f) show pumped current versus device length calculated for a cicular contour of radius equal to the half of the energy level spacing for the corresponding length. The frequency was set to 10 GHz.}
	\label{fig3}
\end{figure}

To rationalize the origin of this dramatic increase in the pumped charge we first observe that from Eq. (\ref{kernel}) one can separate the contributions to the pumping kernel $dQ$ due to the transmission ($dQ_t$) and the reflection amplitudes ($dQ_r$), $dQ=dQ_t+dQ_r$. In our case, where one has low resistance contacts, the latter contribution turns out to be the dominant one (Fig.\ref{fig3} shows the maxima of $dQ$ (red dots) as well as the transmission contribution $dQ_t$ (grey squares)). Indeed, whereas unitariy enforces that a variation in the transmission probability leads to the same (but opposite) variation in the reflection probability, when we analyze the variation in the corresponding phases, the phase of the reflection amplitude is the one that changes in a stronger way. 
In our case, the defect acts as an additional scatterer introducing important changes in the path described by the reflection amplitude in the complex plane as the gate voltages are changed.

Now we turn to another experimentally relevant issue: the scaling of the pumping features with the device width and length. While for the pristine system the maxima of the pumping kernel $dQ^{max}$ remains constant as a function of the device width, it decreases as $1/
\textrm {Width}$ (see Fig. \ref{fig2}-a-b). This is expectable since our pristine model can be reduced to a one-dimensional system with width-independent parameters (by using a mode decomposition as in \cite{Rocha2010}), while the influence of the defect becomes stronger in the 1d limit. On the other hand, $dQ^{max}$ increases with the device length (see Fig. \ref{fig2}-c-d) which scales down the mean level spacing leading to a faster variation of the scattering matrix elements.

To obtain a meaningful value for the pumped current we choose a contour that encircles a single maximum/minimum of $dQ$ in parameter space. Since, the distance between these extrema is locked to the level spacing which scales as the inverse length, our pumping contour becomes smaller as $L^{-2}$. When combined, these two trends give the behavior observed in Fig. \ref{fig2}-e-f for the pumped current as a function of the device length. Notably, in the absence of decoherence effects, the pumped charge for the defected system remains constant even when the chosen contour's area decreases (Fig. \ref{fig2}-f). Furthermore, as shown in Fig. \ref{fig2}-f, the pumped charge surpasses the picoamperes scale for a frequency of 10 GHz.

In summary, we show that defects may dramatically enhance the pumped current in graphene-based devices. Although here we present only the case of a Stone-Wales defect, our simulations show that this effect is generic and does not depend much on the nature of the defect. For instance, a vacancy or a substutional atom give a similar increase in the pumped current. Similar features are also recovered in a purely one-dimensional model for a conductor in the Fabry-P\'erot regime, thereby confirming the generality of the results. While the overall shape of the pumping kernel $dQ$ as a function of the gate voltages changes when adding more defects, the order of magnitude change in the pumped charge is kept in this weak disorder limit. The typical currents obtained could easily reach the pA-nA range for reasonable frequencies $10-100$ GHz. We encourage further experimental work to unveil these exciting phenomena.

We acknowledge the support from SeCyT-UNC, CONICET and ANPCyT through project PICT-PRH 61. We thank P. Orellana, L. Rosales and C. Nu\~nez for useful discussions.



\begin{thebibliography}{10}
\bibitem{footnote}
The parameter controlling the transition from coherent, almost ballistic, transport to a sequential regime dominated by the charging energy is the coupling to the leads, see for example \cite{footnote2,footnote3}.
\end{thebibliography}

%

\end{document}